\documentclass{article}
\usepackage{amssymb}
\usepackage{fleqn}
\usepackage{epsfig}
\usepackage{graphicx}
\usepackage{amsmath}
\def\beq{\begin{equation}}
\def\eeq{\end{equation}}
\def\bea{\begin{eqnarray}}
\def\eea{\end{eqnarray}}
\def\bq{\begin{quote}}
\def\eq{\end{quote}}

\def\gappeq{\mathrel{\rlap {\raise.5ex\hbox{$>$}}
{\lower.5ex\hbox{$\sim$}}}}
\def\lappeq{\mathrel{\rlap{\raise.5ex\hbox{$<$}}
{\lower.5ex\hbox{$\sim$}}}}
\def\Toprel#1\over#2{\mathrel{\mathop{#2}\limits^{#1}}}

\begin{document}

\pagestyle{empty}
\begin{flushright}
ROME1/1377/04~

DSFNA/14/04~~~~~
\end{flushright}
\vspace*{15mm}

\begin{center}
\textbf{A model for next-to-leading order resummed form factors} \\[0pt]

\vspace*{1cm}

\textbf{Ugo Aglietti}\footnote{e-mail address: Ugo.Aglietti@roma1.infn.it} \\[0pt]

\vspace{0.3cm}
Dipartimento di Fisica,\\
Universit\'a di Roma ``La Sapienza'' \\
and I.N.F.N.,
Sezione di Roma, Italy. \\[1pt]

\vspace{0.3cm}\textbf{Giulia Ricciardi}\footnote{e-mail address:
Giulia.Ricciardi@na.infn.it} \\ [0pt]

\vspace{0.3cm} Dipartimento di Scienze Fisiche,\\
Universit\'a di Napoli ``Federico II'' \\
and I.N.F.N.,
Sezione di Napoli, Italy. \\ [1pt]

$~~~$ \\[0pt]
\vspace*{2cm} \textbf{Abstract} \\[0pt]
\end{center}

We present a model for next-to-leading order resummed threshold
form factors based on a time-like coupling recently introduced
in the framework of small $x$ physics.
Improved expressions for the form factors in
$N$-space are obtained which are not plagued by Landau-pole
singularities, as the included absorptive effects
-- usually neglected --- act as regulators.
The physical reason is that, because of faster decay of gluon jets,
there is not enough resolution time to observe the Landau pole.
Our form factors reduce to the standard ones when the absorptive
parts related to the coupling are neglected.
The inverse transform from $N$-space to $x$-space can be done directly
without any prescription and we obtain analytical expressions
for the form factors, which are well defined in all $x$-space.

\vspace*{4cm} \noindent %\rule[.1in]{16.5cm}{.002in}

\vfill\eject
%\pagestyle{empty}
%\clearpage\mbox{}\clearpage

\setcounter{page}{1} \pagestyle{plain}

\section{Introduction}

Resummation of large infrared logarithms in form factors and shape
variables is essential in order to predict accurate cross sections
in many phenomenologically relevant processes
\cite{parpet,kodtren,sterman,cattren,cattren2}. In this paper we
present a model for next-to-leading order (NLO) resummed form
factors based on the time-like coupling recently introduced by B.
Ermolaev, M. Greco and S. Troyan in \cite{ermolaev} in the
framework of small $x$ physics (see also \cite{pennington}). The
usual expression for resummed threshold form factors in $N$-space
is \cite{sterman,cattren,cattren2}:
\begin{eqnarray}
\label{generale}
f_N(\alpha_S) &=& \exp \int_0^1 dz
\frac{z^{N-1}-1}{1-z} \Big\{ \int_{Q^2(1-z)^2}^{Q^2(1-z)}
\frac{dk_t^2}{k_t^2} \big[ A_1\,\alpha_S(k_t^2) +
A_2\,\alpha_S(k_t^2)^2 \,+\,\cdots \big]~+
\nonumber\\
&&~~~~~~~~~~~~~~~~~~~~~~~~~~~
+\,B_1\,\alpha_S(Q^2(1-z))\,+\,\cdots \,+\, D_1\,\alpha_S(Q^2(1-z)^2)+\cdots\Big\}.
\end{eqnarray}
$Q$ is the hard scale of the process.
$A_1,\,A_2,\cdots B_1,\cdots D_1,\cdots$ are the first coefficients  of
the functions $A\left(\alpha_S\right)$, $B\left(\alpha_S\right)$ and
$D\left(\alpha_S\right)$:
\begin{equation}
 A\left(\alpha_S\right) =
  A_{1}\alpha _S+A_{2}\alpha_S^{2}+\cdots, \quad
B\left( \alpha_S\right) = B_1\alpha_S+B_2\alpha_S^2+\cdots,
%~~\mathrm{and}~~
\quad D\left(\alpha_S\right) =D_1\alpha _S+D_2\alpha_S^2+\cdots .
\end{equation}
$A\left(\alpha_S\right)$ describes the emission of partons which
are both soft and collinear, $B\left(\alpha_S\right)$ describes
hard and collinear partons while $D\left(\alpha_S\right)$ describe
partons which are soft and at large angles. The knowledge of the
quantities $A_1$, $A_{2}$, $B_1$ and $D_1$ is needed for
resummation at next-to-leading order (see later for definition).
For instance, in the
case of the thrust distribution we have  \cite{thrust}:
\begin{equation}
A_1 = \frac{2 C_F}{\pi },\qquad
A_2 = \frac{2 C_F}{\pi^2}\left[C_{A}\left( \frac{67}{36}-\frac{\pi^2}{12}\right) -\frac{5}{9}n_F T_R\right],\qquad
B_1 = -\frac{3}{2}\frac{C_F}{\pi }, \qquad
D_1 = 0,
\end{equation}
where $C_A=N_C$, $C_F=(N^2_C -1)/(2 N_C)$, $T_R=1/2$, $N_C=3$ is
the number of colors and $n_F$ is the number of active quark
flavors\footnote{The value of $A_2$ is given in the
$\overline{MS}$ scheme for the coupling constant.}. For heavy
flavor decays we have instead \cite{heavyfla,ug3}:
\begin{equation}
A_1 =\frac{ C_F }{ \pi },\qquad
A_2 =\frac{ C_F }{\pi^2}\left[C_A\left(
\frac{67}{36}-\frac{\pi^2}{12}\right) -\frac{5}{9} n_F T_R\right],\qquad
B_1 =-\frac{3}{4}\frac{C_F}{ \pi }, \qquad
D_1 =-\frac{C_F}{\pi}.
\end{equation}
Generally, one assumes $\alpha_S\ll 1$ and uses truncated
$\alpha_S$ expansions for the functions $A(\alpha_S)$,
$B(\alpha_S)$ and $D(\alpha_S)$ in eq.~(\ref{generale}). Since the
running coupling $\alpha_S\left(k_t^2\right)$ is integrated over
all gluon transverse momenta $k_t$ from the hard scale $Q$ down to
zero, the Landau pole is hit. Therefore $\alpha_S$ diverges inside
the integration region --- it is certainly not small there
--- indicating that a truncated expansion for the $A$, $B$ and $D$
functions is not correct. We have conceptually a breakdown of the
scheme: resummed perturbation theory assumes $\alpha_S$ small at
the beginning, while it ends up with a large $\alpha_S$. A
prescription has to be assigned to give a meaning to the formal
expression (\ref{generale}).

The standard solution is to expand the exponent of
eq.~(\ref{generale}) in a function series of the form
\cite{cattren}:
\begin{equation}
\label{series}
f_N\left(\alpha _S\right)  = \exp \left[
L\,g_1\left(\lambda\right) +\sum_{n=0}^{\infty
}\alpha_S^n\,g_{n+2}\left(\lambda\right) \right] =\exp \left[
L\,g_1\left(\lambda\right) +\,g_2\left(\lambda\right)
+\alpha_S\,g_3\left(\lambda\right) +\cdots \right],
\end{equation}
where
\begin{equation}
\lambda~=~\beta_0~\alpha_S(Q^2)~L,~~~~~~~L~=~\log N
\end{equation}
and $\beta_0 = (11/3\;N_C-2/3\; n_F)/(4\pi)$. The functions
$g_i\left( \lambda \right)$ have a power-series expansion:
\begin{equation}
g_i\left(\lambda \right) =\sum_{n=1}^{\infty }g_{i,n}\lambda^n.
\end{equation}
The effects of the Landau pole in this framework are the following:
\begin{enumerate}
\item the series in eq.~(\ref{series}) is divergent as the higher
order functions have factorially growing coefficients
\cite{ug3,gardi}; \item the functions $g_i(\lambda)$ have branch
cuts starting at $\lambda=1/2$  and going up to infinity. The form
factors are then formally well defined up to a critical value
$N_{crit}\sim Q/\Lambda$ ($\Lambda$ is the QCD scale), above which
they acquire a (completely unphysical) imaginary part.
\end{enumerate}
The first problem is solved by means of a truncation of the series
to its first few terms --- typically two or three --- which is
effectively a prescription of the Landau pole. This is the
so-called fixed logarithmic accuracy, which we will describe later.
The resummation to leading order requires the
knowledge of the function $g_1$, the resummation to
next-to-leading order also requires the knowledge of $g_2$, and so
on.
As far as the second problem is concerned, one simply
restricts himself to a fiducial region in $N$-space below
$N_{crit}$. The inverse Mellin transform from $N$-space to
$x$-space, at next-to-leading $\log(1-x)$ accuracy, is well defined
up to $x_{crit}\sim 1-\Lambda/Q$, above which singularities occur.
However, a form factor formally well defined in the whole
$x$-space and containing all the requested $\log(1-x)$ terms can
be obtained by means of an additional prescription for the contour
integration in $N$-space, the so-called minimal-prescription
\cite{minimalpre}.

Another common solution to this problem involves the renormalon
calculus \cite{gardi}.
The latter uses the inconsistency discussed above to get
information about non-perturbative effects, in the form of
power-suppressed corrections to the cross sections:
\begin{equation}
\delta \sigma \sim c \left(\frac{\Lambda}{Q}\right)^b.
\end{equation}
The exponent $b$ of the power
correction can be computed, but not its coefficient $c$. The
resulting information is therefore of a qualitative kind and,
ultimately, perturbation theory is deprived of predictive power.
In fact, one can substantially modify the spectra coming out of a
renormalon calculation by changing the prescriptions in the Borel
plane, which are completely arbitrary.

Our perspective is different: we want to use resummed
perturbation theory in a
predictive way, by curing the "disease" of usual resummation
formula. Therefore we begin by re-analyzing the derivation of the
standard formula. $\alpha_S(k_t^2)$ occurs in eq.~(\ref{generale})
because we are computing a so-called inclusive-gluon-decay
quantity, in which one does not observe the development of the
jets originating from the gluons emitted by the hard partons. One
then sums over all possible final states, i.e. formally over all
cuts of dressed gluon propagators, reconstructing their
discontinuity. The radiated gluons have a positive virtuality
$k^2>0$, as the development of jets is intrinsically a time-like
process. Absorptive effects, i.e. the $-i\pi$ terms in gluon
polarization functions, are usually neglected in literature:
by taking them into account, one enforces that all
gluons participating to the cascade are unstable particles
\cite{ermolaev}. They are analogous to the quasi-particles of
statistical physics, possessing in lowest order
--- non interacting quasi-particles --- a non-zero width
\cite{quasiparticles}.
The outcome of the inclusion of this physical effect is that
$\alpha_S(k_t^2)$ does not occur any longer in the resummation
formula, and it is replaced by a different function of $k_t^2$,
\begin{equation}
\alpha_S(k_t^2)~\rightarrow~\tilde{\alpha}_S(k_t^2),
\end{equation}
which does not possess the Landau pole. We call this
new function "effective coupling", as it specifies the effective
strength of the interaction in the gluon cascade --- not in a
general QCD process. The main point is that the effective coupling
never becomes large, as the typical expansion parameter is
\begin{equation}
\frac{\tilde{\alpha}_S  (k_t^2) } {\pi} <
\frac{1}{\beta_0\pi} < 1~~~~~~~~{\rm for~~any}~~~k_t^2
\end{equation}
and one never leaves the perturbative domain.
By including absorptive effects related to the coupling constant ---
usually neglected --- we derive an improved expression for the
resummation formula of the form factors. The resummation formula is
free of Landau pole pathologies, it does not involve any new free
parameter and is strictly predictive.

\section{The improved resummation formula}

%
%                     FIGURA COUPLING EFFETTIVO
%
%
\begin{figure}[t]
%% produce figure here
\par
\begin{center}
\epsfxsize=0.40\textwidth \epsfysize=0.25\textheight
\leavevmode\epsffile{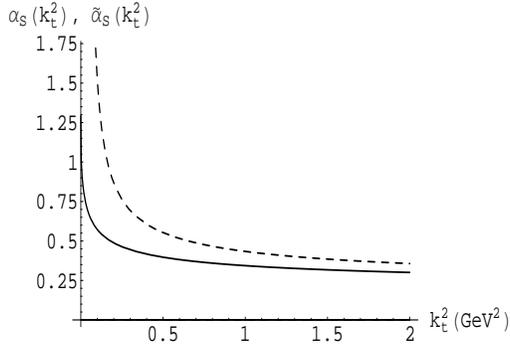}
\end{center}
\caption{\textrm{Coupling constant as a function of gluon
transverse momentum squared $k_t^2$. Solid line: effective
coupling $\tilde{\alpha}_S(k_t^2)$, eq.~(\ref{newcoup}); dashed
line: standard coupling $\alpha_S(k_t^2)$, eq.~(\ref{oldcoup}). }}
\label{fig0}
\end{figure}

By taking into account renormalization effects in the form
factors, the tree-level, momentum-independent coupling is replaced
by the integral on the discontinuity of the gluon propagator
\cite{dokshitzer,amati}:
\begin{equation}
\label{partenza} \alpha_S~~~\rightarrow~~~\tilde{\alpha}_S(k_t^2)
= \frac{1}{2\pi i} \int_0^{k_t^2} ds~\mathrm{Disc}_s ~\frac{1}{
s~\beta_0 \log \frac{-s}{\Lambda^2} }.
\end{equation}
The discontinuity is defined as usual: $\rm{Disc}_s
F(s)=F(s+i\epsilon)-F(s-i\epsilon),$ with $\epsilon$ being a
positive infinitesimal number. We now close the integration
contour with a circle of radius $k_t^2$ and a circle of
infinitesimal radius. By using Cauchy's theorem and neglecting the
residue of the pole in $s=-\Lambda^2$ in order the preserve
asymptotic freedom~\footnote{The integral on the discontinuity of
the gluon propagator is transformed into the integral over a
circle of radius $k_t^2$ {\it plus} the residue of the pole in
$s=-\Lambda^2$. In our pragmatic approach, we have just omitted
the latter.},
we obtain:
\begin{equation}
\tilde{\alpha}_S(k_t^2) = \int_{-\pi}^{+\pi} \frac{d\varphi}{2\pi}
\frac{1}{ \beta_0 \left[ \log \frac{k_t^2}{\Lambda^2} + i\varphi
\right] }. \label{geometric}
\end{equation}
The standard resummation formula (\ref{generale}) is obtained by
neglecting the imaginary term $i\varphi$ in the denominator. In
hard processes $Q\,\gg\,\Lambda$ and one expects on physical
grounds that $k_t\sim Q$, implying:
\begin{equation}
\label{nonvera}
\log \frac{k_t^2}{\Lambda^2} \, \gg \, \pi.
\end{equation}
As a consequence, one obtains the coupling evaluated at the gluon
transverse momentum squared:
\begin{equation}
\tilde{\alpha}_S(k_t^2)~\rightarrow~\frac{1}{\beta_0 \log \frac{k_t^2}{\Lambda^2} }
~=~\alpha_S(k_t^2).
\label{oldcoup}
\end{equation}
However, the assumption (\ref{nonvera}) is not correct, because
the transverse momentum $k_t$ is integrated from $Q$ down to very
small scales. We propose to modify the resummation formula
(\ref{generale}) according to this criticism: approximation
(\ref{nonvera}) disregards absorptive effects related to the gluon
jet decay, which are important in the infrared region, and
therefore has to be avoided. By performing the integration
exactly, we obtain:
\begin{equation}
\label{newcoup}
\tilde{\alpha}_S(k_t^2) =
\frac{1}{\beta_0} \left[ \frac{1}{2} -
\frac{1}{\pi} \arctan \frac{ \log \frac{k_t^2}{\Lambda^2} }{\pi} \right].
\end{equation}
The effective coupling (\ref{newcoup}) approaches the standard one
in the asymptotic region $k_t^2\gg \Lambda^2$, but it does not
contain the infrared pole in $k_t^2 = \Lambda^2$ \footnote{We
define the $\arctan$ function as the one being continuous in zero
and discontinuous at infinity, with image in the interval
$(-\pi/2,+\pi/2)$.}. Furthermore, $\tilde{\alpha}_S(k_t^2)$ is
positive definite, monotonically decreasing in all the $k_t$ range
and has a finite limit at zero momentum (see fig.~1):
\begin{equation}
\lim_{k_t^2\rightarrow 0}\tilde{\alpha}_S(k_t^2) ~=~ \frac{1}{\beta_0}.
\label{zerolimit}
\end{equation}
According to eq.~(\ref{newcoup}), the effective coupling deviates
from the standard one and saturates at a scale of order
\begin{equation}
k_t ~\sim~ \Lambda \, e^{\pi/2} ~\sim~ 1 \; \rm{GeV},
\end{equation}
for $\Lambda\sim 200$ MeV.
%
%
%    SPAZIO N (1)
%
%
\begin{figure}[t]
%% produce figure here
\par
\begin{center}
\epsfxsize=0.40\textwidth \epsfysize=0.25\textheight
\leavevmode\epsffile{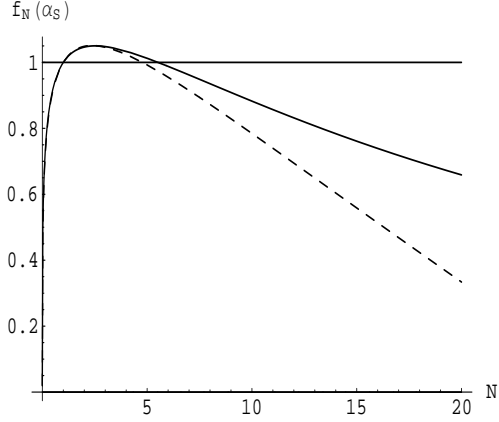}
\end{center}
\caption{\textrm{Form factor in $N$-space $f_N(\alpha_S)$ for the
{\it beauty} case. Solid line: our model; dashed line: standard
form factor, i.e. $r=0$ in our model. } } \label{fig1}
\end{figure}
Our improved expression for the form factor at one-loop
approximation
 reads:
\begin{equation}
\label{generale_new}
f_N(\alpha_S) = \exp \int_0^1 dz \frac{z^{N-1}-1}{1-z}
\Big\{ \int_{Q^2(1-z)^2}^{Q^2(1-z)} \frac{dk_t^2}{k_t^2} \, A_1\,\tilde{\alpha}_S(k_t^2)
\,+\,B_1\,\tilde{\alpha}_S(Q^2(1-z))\,+\, D_1\,\tilde{\alpha}_S(Q^2(1-z)^2)\Big\},
\end{equation}
where the expression for the effective coupling $\tilde{\alpha}_S$
is given by eq.~(\ref{newcoup}). By comparing with the standard
resummation formula (\ref{generale}), we see that the only
difference is the appearance of $\tilde{\alpha}_S$ in place of
$\alpha_S$.
%
%
%    SPAZIO N (2)
%
\begin{figure}[t]
%% produce figure here
\par
\begin{center}
\epsfxsize=0.40\textwidth \epsfysize=0.25\textheight
\leavevmode\epsffile{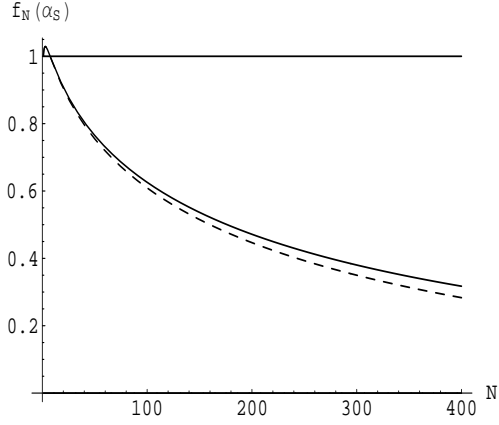}
\end{center}
\noindent \caption{\textrm{$f_N(\alpha_S)$ for the {\it top} case.
Solid line: our model; dashed line: standard form factor, i.e.
$r=0$ in our model.}} \label{fig1b}
\end{figure}
Let us now consider two-loop effects. Note that
eq.~(\ref{partenza}) can be written as:
\begin{equation}
\label{rewrite}
\tilde{\alpha}_S(k_t^2) = \frac{1}{2\pi i} \int_0^{k_t^2} ds~\mathrm{Disc}_s
~\frac{\alpha_S(-s)^{1L}}{ s },
\end{equation}
where
\begin{equation}
\alpha_S(\mu^2)^{1L} = \frac{1}{\beta_0 \log \frac{\mu^2}{~\Lambda^2} }
\end{equation}
is the one-loop coupling. We now assume that eq.~(\ref{rewrite})
generalizes to the two-loop coupling
\begin{equation}
\label{coupling}
\alpha_S\left(\mu^2\right) = \frac{1}{\beta_0\log \mu^2/\Lambda^2}
-\frac{\beta_1}{\beta_0^3}\,\frac{\log
\log \mu^2/\Lambda^2}{\log^2\mu^2/\Lambda^2},
\end{equation}
where $\beta_1 = (51-19/3\;n_F)/(8\pi^2)$. Therefore, the
subleading contribution to the effective coupling
$\tilde{\alpha}_S(k_t^2)$ is:
\begin{eqnarray}
&&  \frac{1}{2\pi i} \int_0^{k_t^2}\,d s \,\mathrm{Disc}_s \,
\frac{1}{s} \,\, \frac{-\beta_1\log \log
(-s/\Lambda^{2})}{\beta_0^3\,\log^2 (-s/\Lambda^2)\,}
\nonumber\\
&=&
-\,\frac{\beta_1}{\beta_0^3} \frac{1}{\pi^2+ \log^2
k_t^2/\Lambda^2}
\left\{ \left[ -\frac{1}{2}  +
\frac{1}{\pi}\arctan \frac{\log k_t^2/\Lambda^2}{\pi } \right] \,
\log \frac{k_t^2}{{\Lambda}^2} + \frac{1}{2} \log \left[ \pi^2 +
\log^2 \frac{k_t^2}{\Lambda^2} \right] + 1 \right\}.
\end{eqnarray}
It is not easy to derive the terms proportional to $A_2$.
Let us make the following guess:
\begin{eqnarray}
\label{coupling_doppio}
\alpha_S^2(k_t^2)~\rightarrow~\tilde{\alpha_S^2}(k_t^2)\,&=&\,
 \frac{1}{2\pi i} \int_0^{k_t^2}d s \, \mathrm{Disc}_s
~\frac{\alpha^2_S(-s)^{1L}}{ s }
\nonumber \\
&=&
\frac{1}{2\pi i} \int_0^{k_t^2}d s \, \mathrm{Disc}_s
\frac{1}{s~\beta_0^2\log^2 (-s/\Lambda^2) }
\nonumber \\
&=& \frac{1}{\beta_0^2(\pi^2+
\log^2 k_t^2/\Lambda^2)}.
\end{eqnarray}
An explicit verification of the validity of
eq.~(\ref{coupling_doppio}) would require a three loops
computation, which is beyond the purpose of the present work. That
is because $A_2$ first occurs at order $\alpha_S^2$ and running
coupling effects involve at least one additional power of
$\alpha_S$, i.e. $\alpha_S^3$ in total. Let us stress that also
the standard replacement $\alpha_S^2 \rightarrow \alpha_S(k_t)^2$
has never been explicitly checked, as its verification requires a
three loop computation as well. In eq.~(\ref{coupling_doppio}) we
just assumed that the $s$-discontinuity had to be taken after
squaring the gluon polarization function; it is remarkable that
our choice gives a simple analytic result. Other choices are
possible at this level, such as squaring the effective coupling:
$\alpha_S^2(k_t^2) \rightarrow \tilde{\alpha}_S(k_t^2)^2$
\footnote{With this choice there is only one effective coupling,
while with our choice there are two effective couplings.}. Anyway,
since two-loop terms are rather small, our choice seems acceptable
in a model of soft gluon dynamics.

In conclusion, our expression for the resummed form factor at
next-to-leading order reads:
\begin{eqnarray}
\label{improved} f_N(\alpha_S) &=& \exp \int_0^1 dz
\frac{z^{N-1}-1}{1-z} \Big\{ \int_{Q^2(1-z)^2}^{Q^2(1-z)}
\frac{dk_t^2}{k_t^2} \big[ A_1 \tilde{\alpha}_S(k_t^2) +  A_2
\tilde{\alpha_S^2}(k_t^2) + \cdots \big]\,+
\nonumber\\
&&~~~~~~~~~~~~~~~~~~~~~~~~~~~
+\,B_1 \tilde{\alpha}_S(Q^2(1-z))+\cdots \,
+\, D_1 \tilde{\alpha}_S(Q^2(1-z)^2)+\cdots\Big\},
\label{espansione}
\end{eqnarray}
where:
\begin{eqnarray}
\label{alphaS2L}
\tilde{\alpha}_S(k_t^2)&=&
\frac{1}{\beta_0} \left[ \frac{1}{2} -
\frac{1}{\pi} \arctan \frac{ \log \frac{k_t^2}{\Lambda^2} }{\pi} \right] +
\nonumber\\
&-&\frac{\beta_1}{\beta_0^3}
\frac{1}{\pi^2+ \log^2 k_t^2/\Lambda^2}
\left\{\left[ -\frac{1}{2}  +
\frac{1}{\pi}\arctan \frac{\log k_t^2/\Lambda^2}{\pi } \right] \,
\log \frac{k_t^2}{{\Lambda}^2} + \frac{1}{2} \log \left[ \pi^2 +
\log^2 \frac{k_t^2}{\Lambda^2} \right] + 1 \right\}
\end{eqnarray}
and $\tilde{\alpha_S^2}(k_t^2)$ is given
by eq.~(\ref{coupling_doppio}).

Eq.~(\ref{improved}) is the main result of this work and replaces
eq.~(\ref{generale}). The main problem of eq.~(\ref{generale}),
discussed before, does not occur anymore in our improved
expression: the effective couplings $\tilde{\alpha}_S$ and
$\tilde{\alpha_S^2}$ do not diverge for small values of the
argument and actually remain acceptably small.

Our treatment of the running coupling constant resembles  the
exponentiation of the $\pi^2$ terms in the $K$-factor of Drell-Yan
coefficient function \cite{prima}. It is well known that, going
from the space-like kinematics of Deep-Inelastic-Scattering
($q^2<0$) to the time-like kinematics of Drell-Yan ($q^2>0$),
double logarithms in the vertex correction generate a large
constant term proportional to $\pi^2$, because:
\begin{equation}
\log^2(-q^2-i0) \,\rightarrow\, \log^2(q^2) - \pi^2 + \cdots.
\end{equation}
Since double logs exponentiate, it has been suggested that
also the $\pi^2$ terms exponentiate:
\begin{equation}
e^{ -c\,\alpha_S\,\log^2 (- q^2 - i 0) } \, \rightarrow \,
e^{ c\,\alpha_S\,\pi^2 } \, e^{ -c\,\alpha_S\,\log^2 (q^2) },
\end{equation}
where $c$ is a positive constant \footnote{In our work we are
dealing with a {\it geometric} series (cfr. eq.
(\ref{geometric})), while in the $K$-factor an {\it exponential}
series is resummed.}. It has been however established that the
exponentiation of $\pi^2$ is not exact, being violated  at two
loops  by terms of the form $\zeta(2)\log^2(-q^2-i0)$, where
$\zeta$ is the Rieman zeta function ($\zeta(2)=\pi^2/6$)
\cite{prima2}. Analogously, in our case, we cannot expect to have
a complete resummation; we resum, however, a set of constants
which are certainly there, and this seems to be a reasonable
approximation.

\section{The functions $g_1$ and $g_2$}

Let us begin this section by defining the logarithmic accuracy in
our model. We have powers of the coupling at the hard scale,
\begin{equation}
\alpha\equiv\alpha_S(Q^2),
\end{equation}
multiplied by powers of the infrared logarithm, $L=\log N$, i.e.
monomials of the form
\begin{equation}
\alpha^n ~ L^k.
\end{equation}
Within the usual resummation scheme,
\begin{itemize}
\item
by leading logarithms ($LL$) we mean all the terms
in the exponent of the form factor of the form
\begin{equation}
\alpha^n ~ L^{n+1} ~~~~~~~~~~~~~~ {\rm with}~~~n = 0, 1,2,\cdots\infty ;
\end{equation}
\item
by next-to-leading logarithms ($NL$) we mean all the terms of the form
\begin{equation}
\alpha^n ~ L^n     ~~~~~~~~~~~~~~~ {\rm with}~~~n = 0, 1,2,\cdots\infty.
\end{equation}
\end{itemize}
The coupling $\alpha_S(k_t^2)$ at one-loop, when expanded in
powers of $\alpha$, produces a series of leading logarithms:
\begin{equation}
\label{expand1}
\alpha_S(k_t^2) = \alpha - \beta_0\alpha^2\log\frac{k_t^2}{Q^2}
+ \beta_0^2\alpha^3\log^2\frac{k_t^2}{Q^2}
+ \cdots
\end{equation}
The powers of $\log k_t^2/Q^2$ give indeed  powers of
$L=\log N$ after the integration over $k_t$ and $z$ in eq.
(\ref{generale}).

According to eq.~(\ref{newcoup}), the relation between
the standard and the effective coupling reads
\footnote{One has to use the relation $\arctan(1/x) = \pi/2-\arctan(x)$,
which is valid for $x\geq0$ with our definition of the $\arctan$
function.}:
\begin{equation}
\tilde{\alpha}_S(k_t^2) =
\frac{1}{\pi\beta_0}\arctan\left[\pi\beta_0\alpha_S(k_t^2)\right]
~~~~~{\rm for}~~~\alpha_S(k_t^2) > 0.
\end{equation}
Therefore, the expansion of our effective coupling in the parameter $\alpha$ is:
\begin{equation}
\label{expand2}
\tilde{\alpha}_S(k_t^2) = \alpha - \beta_0\alpha^2\log\frac{k_t^2}{Q^2}
+ \beta_0^2\alpha^3\log^2\frac{k_t^2}{Q^2}
- \frac{(\beta_0\pi)^2}{3} \alpha^3 + \cdots
\end{equation}
Comparing eq.~(\ref{expand1}) with eq.~(\ref{expand2}), we see
that the expansion of the effective coupling involves subleading
logarithms compensated by powers of $\beta_0\pi$, coming from the
absorptive effects discussed before. The main point is that the
$\beta_0\pi$ terms have a fundamental regulating effect over the
logarithmic ones; therefore one has to consider them on equal
footing in the counting. The hierarchy in our model can therefore
by defined as follows:
\begin{itemize}
\item
by $LL$ we mean all the terms of the form
\begin{equation}
\alpha^n ~ (\beta_0 L)^{n+1-k} ~ (\beta_0\pi)^k ~~~{\rm with}~~~ k
= 0, 1,\cdots n+1;
\end{equation}
\item by $NL$ we mean all the terms of the form
\begin{equation}
\alpha^n ~ (\beta_0 L)^{n-k} ~ (\beta_0\pi)^k ~~~{\rm with}~~~k =
0, 1,\cdots n.
\end{equation}
\end{itemize}
The usual scheme is ``minimal'' in the sense that it  deals only
with powers of the logarithms and of the coupling; the related
problem is that of the Landau pole discussed above. Instead, our
scheme, in order to solve the Landau problem, has to be
non-minimal.

The next step is to perform the integrations within the above
defined next-to-leading accuracy. In general, we believe that
fixed logarithmic accuracy is a consistent approximation in our
scheme because series of logarithms of higher order are suppressed
by powers of $\tilde{\alpha}_S$, which is always small. The
integrations over transverse and longitudinal gluon momenta are
easily done within the standard approximation \cite{cattren}:
$z^{N-1}-1\simeq -\theta \left(1-z-1/n\right),\label{ctapprox} $
where $\theta$ is the step function and $n=N e^{\gamma_E}$, with
$\gamma_E$ being the Euler constant. This approximation, as shown
in our previous paper \cite{ug3}, misses for instance terms of the
form: $ \pi^2 \alpha^n L^{n-1}$, which are leading according to
our power counting. However, in our model, we are interested in
picking up only those terms proportional to $\beta_0\pi$, which
come from the analytic continuation of the coupling from
space-like to time-like region. These are the terms coming from
the effective coupling, having the regulating effect on the Landau
pole.

In order to obtain the functions $g_1$ and
$g_2$, we express the logarithm of the hard scale $\log
Q^2/\Lambda^2$, coming from the integrations, in terms of the
coupling evaluated at a general renormalization scale $\mu$ by
means of the formulas:
\begin{equation}
\log \frac{Q^2}{\Lambda^2} = \log\frac{\mu^2}{\Lambda^2} - \log \frac{\mu^2}{Q^2}.
\end{equation}
and
\begin{equation}
\log \frac{\mu^2}{\Lambda^2} =\frac{1}{\beta _{0}\alpha _S(\mu^2) }
+\frac{\beta_1}{\beta_0^2} \log \left(\beta_0\, \alpha _S(\mu^2)
\right).
  \label{logsrep}
\end{equation}
We finally Taylor expand the exponent on the r.h.s of
eq.~(\ref{generale_new}) in powers of $\gamma_E$, $\beta_1$ and
$\log\mu^2/Q^2$ up to first order included. Our result for the
leading function $g_1$ reads:
\begin{eqnarray}
\label{g1new}
& & g_1(\lambda;r) = \frac{\mathrm{A_1}}{4\,\beta_0\, \lambda}
    \biggl\{ -\left( 1 - 2\,\lambda \right) \,
       \log [(1 - 2\,\lambda)^2+r^2]
     + 2\,  \left( 1 - \lambda \right)
           \,\log [(1 - \lambda)^2+r^2]- \log[1+r^2]+
\\
& & + \left. \frac{1}{r}
 \left[ -\pi\,\lambda^2 + (1-r^2)\,\arctan \frac{1}{r}+
\left[ \left( 1 - 2\lambda  \right)^2-r^2 \right]
 \arctan \frac{1-2\lambda}{r} -
2 \left[\left( 1 - \lambda \right)^2-r^2\right]
 \arctan \frac{1-\lambda}{r}
           \right]
       \right\},
\nonumber
\end{eqnarray}

The subleading function $g_2$ reads:
\begin{eqnarray}
\label{g2new}
g_2(\lambda;r) &=& \frac{\mathrm{D_1}}{ 4 \beta_0}
    \left\{
       \log [(1 - 2\,\lambda)^2+r^2]-\log [1+r^2]
+ \frac{2}{r}
 \left[
- \pi\,\lambda+ \arctan \frac{1}{r} - ( 1 - 2\lambda )  \arctan
\frac{1- 2 \lambda}{r}\right] \right\}+
     \nonumber \\
&+& \frac{\mathrm{B_1}}{2\,\beta_0}
    \left\{
       \log [(1 - \lambda)^2+r^2]-\log [1+r^2]
+ \frac{1}{r} \left[-\pi\,\lambda+
 2\,\arctan\frac{1}{r} -2\, ( 1 -
\lambda )  \arctan \frac{1-\lambda}{r}\right] \right\} + \nonumber
\\
&+& \frac{\mathrm{A_2}} {4\, \beta_0^2} \biggl\{ \log[(1- 2
\lambda)^2 + r^2] -2 \log[(1- \lambda)^2 + r^2] + \log[1+ r^2]+
\nonumber \\
&+ & \left. \frac{2}{r} \left[- ( 1- 2 \lambda) \arctan\frac{1-2
\lambda}{r}+ 2 ( 1-  \lambda) \arctan\frac{1-\lambda}{r} -
 \arctan\frac{1}{r}\right]
 \right\} + \nonumber \\
&+&
 \frac{\mathrm{A_1 \,\beta_1}}{ 16\, \beta_0^3} \biggl\{
-\log^2[(1- 2 \lambda)^2 + r^2] + 2  \log^2[(1- \lambda)^2 +
r^2]-\log^2[1+ r^2] +
\nonumber \\
&+& \left. \frac{2\,\pi}{r} ( 1-2 \lambda) \left[-1+\frac{2}{\pi}
\arctan\frac{1-2\lambda}{r}\right] \log[(1-2\lambda)^2+r^2] +
 \nonumber \right. \\
&+& \left.
 \frac{4\,\pi}{r} ( 1- \lambda) \left[1-\frac{2}{\pi}
\arctan\frac{1-\lambda}{r}\right] \log[(1-\lambda)^2+r^2]+
\frac{2\,\pi}{r}  \left[-1+\frac{2}{\pi} \arctan\frac{1}{r}\right]
\log[1+r^2] +
 \nonumber \right. \\
&+& \left. 4 \arctan \frac{1-2\lambda}{r}\left[-\pi+
\arctan\frac{1-2\lambda}{r}\right]+ 8 \arctan
\frac{1-\lambda}{r}\left[\pi- \arctan\frac{1-\lambda}{r}\right]+
\right. \nonumber \\
&+& \left. 4 \arctan \frac{1}{r}\left[-\pi+
\arctan\frac{1}{r}\right] \right\}
+ \nonumber \\
&+& \frac{\mathrm{A_1}}{ 4\, \beta_0}\,\log\frac{\mu^2}{Q^2}
\biggl\{ \log[(1- 2 \lambda)^2 + r^2] -2 \log[(1-  \lambda)^2 +
r^2] + \log[1+ r^2]+
 \nonumber \\
&+& \left. \frac{2}{r} \left[- ( 1- 2 \lambda) \arctan\frac{1-2
\lambda}{r}+ 2 ( 1-  \lambda) \arctan\frac{1-\lambda}{r} -
 \arctan\frac{1}{r}\right]
 \right\} + \nonumber \\
  & + &
 \frac{\mathrm{A_1\, \gamma_E}}{ 2\, \beta_0}
\biggl\{ \log[(1- 2 \lambda)^2 + r^2] - \log[(1-  \lambda)^2 +
r^2] +  \nonumber \\ &+ &  \left. \frac{2}{r} \left[- ( 1- 2
\lambda) \arctan\frac{1-2 \lambda}{r}+
 ( 1-  \lambda) \arctan\frac{1-\lambda}{r}
-\frac{\pi \,\lambda}{2}  \right] \right\}.
\end{eqnarray}
We have defined --- note the general renormalization scale $\mu$ in the coupling:
\begin{equation}
\label{lambdardef} \lambda~\equiv~\beta _0\, \alpha _S (\mu^2) \,
L ~~~~~\mathrm{and}~~~~~ r~\equiv~\pi~\beta _{0} \, \alpha
_S(\mu^2).
\end{equation}
Note that $\lambda$ and $r$ are obtained from each other by
interchanging $L$ with $\pi$. All this is in line with the
logarithmic accuracy defined above.

A few remarks are in order. The first one is that, unlike the
standard case, $g_1$ and $g_2$ do not depend only on $\lambda$ but
also on $r$, i.e. on $\alpha_S$ alone. Our $g_1$ and $g_2$ are
then ``non-minimal'', as a consequence of the inclusion of
absorptive constants in the coupling. The parameter $r\neq 0$ acts
as a regulator of Landau-pole singularities: $g_1$ and $g_2$ are
regular for any value of $\lambda$. As expected, in the limit
$r\rightarrow 0$ we recover the standard $g_1$ and $g_2$
\cite{cattren}:
\begin{equation}
\label{g1old}
g_1(\lambda;r=0) = \frac{\mathrm{A_1}}{2\,\beta_0\, \lambda}
    \left\{ -\left( 1 - 2\,\lambda \right) \,
       \log [1 - 2\,\lambda]
     + 2\,  \left( 1 - \lambda \right)
           \,\log [1 - \lambda]  \right\}
\end{equation}
and
\begin{eqnarray}
g_2(\lambda;r=0)&=&
\frac{\mathrm{D_1}\,\log (1 - 2\,\lambda )}
   {2\,{{\beta }_0}} +
\frac{\mathrm{B_1}\,\log (1 - \lambda )}
   {{{\beta }_0}}
+\frac{\mathrm{A_2}\,
     \left[ \log (1 - 2\,\lambda ) -
       2\,\log (1 - \lambda ) \right] }{2\,
     {{{\beta }_0}}^2} +
\\ \nonumber
&-&\frac{\mathrm{A_1}\, {{\beta }_1}\,
     \left[ 2\,\log (1 - 2\,\lambda ) +
       {\log^2 (1 - 2\,\lambda )} -
       4\,\log (1 - \lambda ) -
       2\,{\log^2 (1 - \lambda )} \right] \,
     }{4\,{{{\beta }_0}}^3} +
\\ \nonumber
&+&\frac{\mathrm{A_1}\, \log (\mu^2/Q^2) \,
     \left[ \log (1 - 2\,\lambda ) -
       2\,\log (1 - \lambda ) \right] \,}{2\,{{\beta }_0}}
 + \frac{\mathrm{A_1}\,{{\gamma }_E}\,
     \left[ \log (1 - 2\,\lambda ) -
       \log (1 - \lambda ) \right] }
     {{{\beta }_0}}.
\end{eqnarray}
If we expand our $g_1$ and $g_2$ for small $r$, we find that the
leading corrections are of order $r^2 \lambda^n$: this means
corrections of next-to-next-to-leading order (NNLO)
$\alpha_S^{n+1} L^n$ in the standard counting, which had been
overlooked in our previous evaluation of the $g_3$ \cite{ug3}. Our
$g_1$ and $g_2$ contain some terms from the standard $g_3$, $g_4$,
$g_5$, $\ldots$ or from the coefficient function $C(\alpha_S)$
multiplying the form factor. It is remarkable that
eqs.~(\ref{g1new}) and (\ref{g2new}) involve only powers of the
$\mathrm{log}$ and $\mathrm{arctan}$ functions.

%
%
%    SPAZIO N (3)
%
%
\begin{figure}[t]
% produce figure here
\par
\begin{center}
\epsfxsize=0.40\textwidth \epsfysize=0.25\textheight
\leavevmode\epsffile{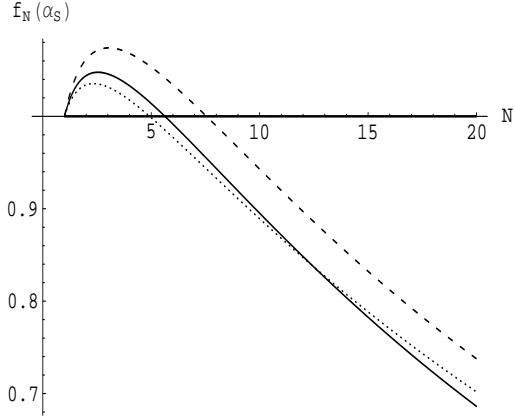}
\end{center}
\caption{\textrm{Renormalization scale dependence of form factor
in $N$-space $f_N(\alpha_S)$ for the {\it beauty} case. Solid
line: $\mu^2=Q^2$; dashed line: $\mu^2=Q^2/4$; dotted line:
$\mu^2=4Q^2$.}} \label{fig2}
\end{figure}
%
%
%
%
%
%
%
%                         x SPACE (1)
%
%
\begin{figure}[t]
%% produce figure here
\par
\begin{center}
\epsfxsize=0.40\textwidth \epsfysize=0.25\textheight
\leavevmode\epsffile{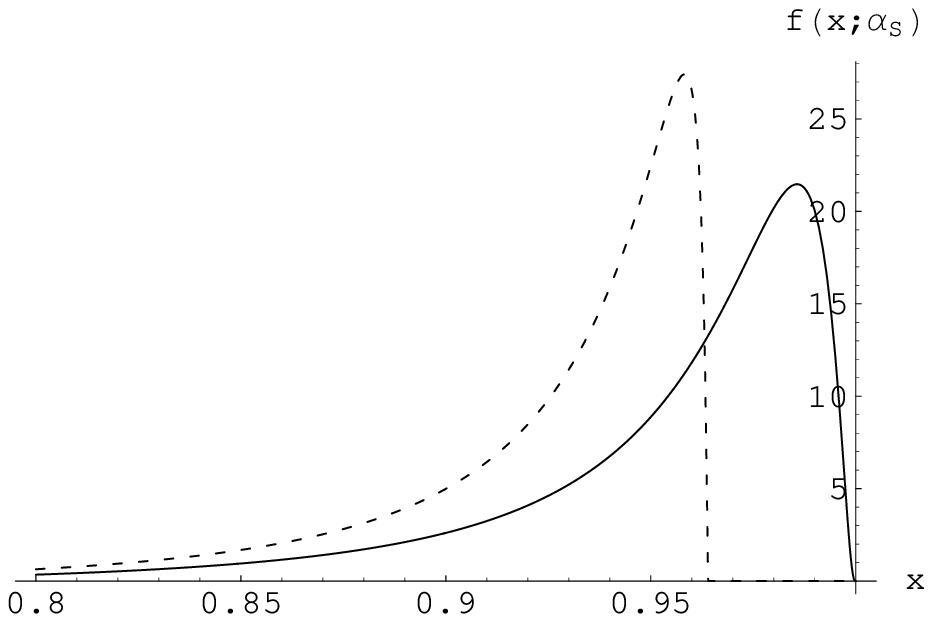}
\end{center}
\caption{\textrm{Form factor in $x$-space $f(x;\alpha_S)$ for the
{\it beauty} case. Solid line: our model; dashed line: standard
form factor.}} \label{fig3}
\end{figure}
In fig.~2 we have plotted the form factor $f_N(\alpha_S)$ as a
function of $N$ for the decay of a beauty quark, i.e. for
$Q=m_b$\footnote{In general, for heavy flavor decays $Q=2E_X$,
where $E_X$ is the energy of the hadronic final state. One can set
$2E_X=m_b$ in $b\rightarrow s\gamma$, while this is not true in
semi-leptonic $b\rightarrow u$ decays \cite{tripledis}.}, for
which $\alpha_S=0.21$ and $n_F=3$. The dashed line is the plot of
the standard form factor, i.e. of the same form factor in the
limit $r \rightarrow 0$. We have plotted moments up to $N=20$,
because above this point the standard form factor, as already
discussed, becomes singular. Our form factor is reasonably close
to the standard one up to  $N\sim 10$, the latter being steeper
for larger $N$. For $N=20$ there is a difference of a factor 2
circa. Fig.~3 shows similar plots for the top case, i.e. for
$Q=m_t$ for which $\alpha_S=0.11$ and $n_F=5$. Due to the
increased hard scale by more than one order of magnitude:
\begin{itemize}
\item differences between the two models are much smaller; \item
the peak above one, related to the subleading, single-logarithmic
effects, is barely visible, while it is rather pronounced in the
{\it beauty} case.
\end{itemize}
The scale dependence is shown in fig.~4, where we have plotted
$f_N(\alpha_S)$ for the {\it beauty} case for $\mu^2=Q^2$,
$\mu^2=Q^2/4$ and $\mu^2=4Q^2$. The scale ambiguity is smaller
than the difference between the models shown in fig.~2.
%
%
%     x SPACE (2)
%
%
\begin{figure}[t]
%% produce figure here
\par
\begin{center}
\epsfxsize=0.40\textwidth \epsfysize=0.25\textheight
\leavevmode\epsffile{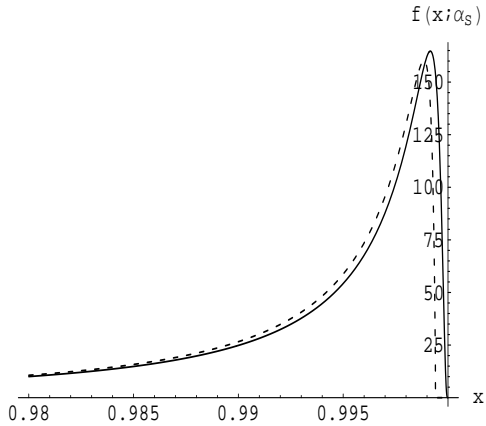}
\end{center}
\caption{\textrm{Form factor in $x$-space $f(x;\alpha_S)$ for the
{\it top} case. Solid line: our model; dashed line: standard form
factor.}} \label{fig4}
\end{figure}

Let us now discuss the extension of our model to NNLO, i.e.
the computation of the function $g_3$. It is clear
that one has to make similar guesses to the ones  needed at NLO
to evaluate the terms proportional to $A_2$. Apart from
that, the computation, although technically rather cumbersome,
does not seem to present any specific difficulty. In general, our
fixed logarithmic accuracy for the exponent in the form factor,
\begin{equation}
\Sigma = L ~ g_1(\lambda;r) + g_2(\lambda;r)
       + \alpha ~ g_3(\lambda;r) + \alpha^2 g_4(\lambda;r) + \cdots ,
\end{equation}
is an expansion with better convergence properties than the usual
one, because the effective coupling becomes at the most of order
one. In other words, our expansion is certainly better than an
expansion in $1/(\beta_0\pi)^n$, which is already convergent as
$1/(\beta_0\pi) < 1$. We expect the extension of our model to NNLO
to be relevant also for beauty physics. In a previous work
\cite{ug3} we have found instead that usual NNLO effects could
not be included in the case of beauty decay. In fact, the hard
scale $Q = m_b$ was not large enough to avoid effects of the
divergence of the perturbative series even at relatively low
values of $x$. Such divergence was related to the integration over
the Landau pole and is therefore absent in the present scheme.

\section{Form Factors in $x$ space}

Up to now we have considered form factors in $N$-space. The
$N$-moments are indeed physical quantities, but in practice a
measure of the moments for large $N$ is difficult. Therefore, let
us transform back to $x$-space. The inverse transform of
$f_N(\alpha_S)$ is defined as:
\begin{equation}
\label{xspace}
f(x;\alpha_S) = \frac{1}{2 \pi i } \int_{C-i \infty}^{C+i
\infty}dN \,x^{-N} f_N(\alpha_S),
\end{equation}
where $C$ is a constant such that the integration contour lies to
the right of all the singularities of $f_N$.
A standard computation gives to next-to-leading $\log(1-x)$ accuracy
\cite{thrust}:
\begin{equation}
\label{finx} f(x;\alpha_S) = -\frac{d}{dx} \left\{ \theta(1-x-0) \frac{ \exp
\left[\,l\, g_1(\tau;r)+g_2(\tau;r) \right] } {
\Gamma\left[1-\frac{d}{d\tau}
 \left(\tau g_1(\tau;r) \right)\right] }  \right\}  ,
\end{equation}
where we have defined:
\begin{equation}
\tau~\equiv~\beta_0~\alpha_S(\mu^2)~l~~~~~\mathrm{and}~~~~~l~\equiv~-\log(1-x).
\end{equation}
In eq.~(\ref{finx}) we have neglected terms $O(1-x)$, which are
small in the threshold region. The form factor $f(x;\alpha_S)$ in
eq.~(\ref{finx}) is plotted in fig.~5 for the {\it beauty} case.
Also shown is the standard form factor, i.e. the same formula for
$r=0$. Our form factor is formally well defined up to $x=1$, while
the standard one presents a singularity at $x_{crit}\sim
1-\Lambda/Q$, so it is plotted only below this point. There are
substantial differences between the models: our distribution is
broader, the peak is smaller and occurs at much larger $x$ than in
the standard case. The reason is the following. Our model contains
an effective coupling which approaches a constant value for
$N\rightarrow\infty$ or, equivalently, $x\rightarrow 1$ (see
eq.~(\ref{zerolimit})). Therefore, for $x\sim 1$ our model
resembles a frozen coupling scheme, while the standard form factor
contain a coupling divergent at the non-zero momentum $\Lambda$
--- the well-known Landau pole. In fig.~6 we plot $f(x;\alpha_S)$
for the {\it top} case in both models; as expected, differences
are much smaller with respect to the {\it beauty} case. The
dependence of our $f(x;\alpha_S)$ from the renormalization scale
$\mu$ is shown in fig.~7 for the {\it beauty} case and turns out
to be moderate.

%
%     x SPACE (3)
%
\begin{figure}[t]
%% produce figure here
\par
\begin{center}
\epsfxsize=0.40\textwidth \epsfysize=0.25\textheight
\leavevmode\epsffile{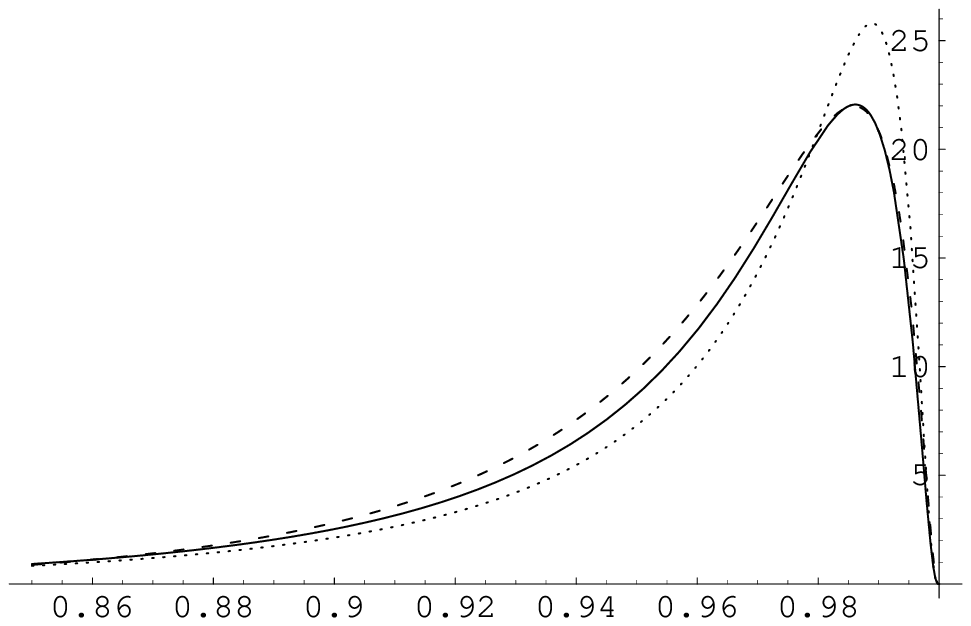}
\end{center}
\caption{\textrm{ Renormalization scale dependence of form factor
in $x$-space $f(x;\alpha_S)$ for the beauty case. Solid line:
$\mu^2=Q^2$; dashed line: $\mu^2=Q^2/4$ ; dotted line: $\mu^2= 4
Q^2$.}}
\label{fig5}
\end{figure}

\section{Phenomenology}

Let us now discuss the phenomenological relevance of our model. A
natural application concerns beauty physics, i.e. a case with a
moderately large hard scale --- when the hard scale is very large,
as in top decays, there are basically no differences between our
model and the standard one. In the beauty case, large
non-perturbative corrections are expected. The main problem, in
our opinion, is that of combining perturbative and
non-perturbative effects in the most efficient way. In many
phenomenological studies, one assumes a shape for the
non-perturbative effects and convolutes it with a frozen-coupling
distribution. Since $\alpha_S(m_b)\sim0.21$ (that is, rather
small) and the coupling does not increase at small $k_t$, the
perturbative effects turn out to be extremely small. Since the
theoretical prediction is a convolution of perturbative and
non-perturbative effects, it is completely dominated by the
assumed form of the latter. The standard resummation scheme
--- running coupling ---
includes much more perturbative dynamics, but it is technically
more difficult to implement in a phenomenological analysis and it
has been barely used. That is because the perturbative form
factors in $x$-space are generally computed with a numerical
contour integration, by using the so-called minimal prescription
\cite{minimalpre}. This is a consistent prescription based again
on "minimality": the correct logarithms of $1-x$ are included and
the distribution is formally well defined up to $x = 1$, but no
physically motivated effects are included, only logarithms.
Our
approach includes instead a (perturbative) physical effect, namely
parton decay, giving rise to analytic distributions well-defined
in the whole $x$-space. One can then easily convolve our spectra
with an assumed form of non-perturbative effects:
\begin{eqnarray}
f_{ex}(x) &=& \int_0^1 \int_0^1 dx' dx''
\delta(x-x'x'') f_{np}(x';c_k) f(x'';\alpha_S)
\nonumber\\
&=&
\int_x^1 \frac{dx'}{x'} f_{np}(x';c_k) f(x/x';\alpha_S),
\end{eqnarray}
where $f_{np}$ is a non-perturbative function and $f_{ex}$
is the spetrum to be compared with the data. $c_k$ denotes
schematically non-perturbative parameters entering $f_{np}$.

It is well known that non-perturbative effects broaden the
peaks of perturbative spectra and shift them to smaller $x$'s.
They have, as intuitively expected, a smearing effect,
representing the final stage of parton evolution, not described
by perturbation theory.
We have not done yet any detailed phenomenological analysis,
but the following points can already be made.
Our model gives a broader spectrum compared to the usual one,
while it gives a peak at larger values of $x$.
The peak we find corresponds indeed to a jet invariant mass below 1 GeV.
We then expect that our model has to be supplemented with
non-perturbative effects shifting the peak towards smaller $x$.

\section{Conclusions}

Our main result is the improved threshold resummation formula
given by eq.~(\ref{improved}).
The latter involves ---
in place of the usual running coupling possessing the Landau pole ---
an effective coupling which is not singular in the infrared
region and remains rather small in all the integration domain.
It approaches the standard coupling for large transverse momenta,
while it deviates from the latter at a scale of order
\begin{equation}
k_t~\sim~\Lambda e^{\pi/2}\sim~1~{\rm GeV}.
\end{equation}
The effective coupling, unlike the usual one, includes absorptive
effects related to the decay of gluon jets. The physical
explanation of the absence of the Landau pole in our resummation
formula is the following. The inclusive gluon branching, inducing
the running coupling in resummation formula, is described by
taking into account absorptive effects in addition to the usual
dispersive ones. The gluon cascade is then treated, more
realistically, as an intrinsically unstable process and there is
not enough resolution time to see the effects of the Landau pole.

By performing the integrations over longitudinal and transverse
momenta with next-to-leading accuracy, we have obtained the
functions $g_1$ and $g_2$ in the exponent of the form factor given
in eqs.~(\ref{g1new}) and (\ref{g2new}) respectively. These
functions define next-to-leading order resummed threshold
distributions which are well defined in the whole moment or
physical space. There are large differences between our
next-to-leading order form factor and the standard one for beauty
decays. We believe that our model can be used to describe the
variety of spectra in inclusive or rare $b$ decays. Natural
extensions of our approach involve resummation of shape variables
or the inclusion of mass effects in semi-leptonic $b\rightarrow c$
transitions.

%\begin{center}
%\bigskip Acknowledgements
%\end{center}

\vskip 0.5truecm
\centerline{\bf Acknowledgments}

One of us (U.A.) wishes to thank B. Ermolaev for discussions.

\end{document}